\documentclass{aa}  

\usepackage{graphicx}
\usepackage{txfonts}
\usepackage[normalem]{ulem}
\usepackage{color}
\usepackage{url}
\usepackage{hyperref} 
\usepackage{caption}
\usepackage{subcaption}
\usepackage{sidecap}
\usepackage{orcidlink}
\usepackage{xspace}

\usepackage[switch]{linenoaa} 
\linenumbers

\newcommand{\lsv}{\mbox{LS~V~+44~17}\xspace}

\begin{document}

\title{Optical linear polarization of the Be--X-ray pulsar binary \lsv}

\titlerunning{Optical polarization of \lsv}

\author{Anagha P. Nitindala\inst{1}\orcidlink{0009-0002-7109-0202}
\corrauth{anaghapradeep.a.nitindala@utu.fi}
\and
Andrei V. Berdyugin\inst{1}\orcidlink{0000-0002-9353-5164}
\and
Alexandra Veledina\inst{1,2}\orcidlink{0000-0002-5767-7253}
\and
Sergey S. Tsygankov\inst{1}\orcidlink{0000-0002-9679-0793}
\and
Vadim~Kravtsov\inst{3,1}\orcidlink{0000-0002-7502-3173}
\and Mar\'ia~Alejandra~D\'iaz~Teodori\inst{1}\orcidlink{0009-0002-1852-7671}
\and Vilppu Piirola\inst{1}\orcidlink{0000-0003-0186-206X}
\and Takeshi Sakanoi\inst{4}\orcidlink{0000-0002-7146-9020} 
\and Masato~Kagitani\inst{4}
\and Svetlana~V.~Berdyugina\inst{5}\orcidlink{0000-0002-2238-7416}
\and
Juri Poutanen\inst{1}\orcidlink{0000-0002-0983-0049}
}

\institute{Department of Physics and Astronomy, FI-20014 University of Turku, Finland
%\email{anaghapradeep.a.nitindala@utu.fi}
\and
Nordita, Stockholm University and KTH Royal Institute of Technology, Hannes Alfv\'ens v\"ag 12, SE-10691 Stockholm, Sweden
\and
Department of Physics, P.O. Box 64, FI-00014  University of Helsinki, Finland
\and Graduate School of Sciences, Tohoku University, Aoba-ku, 980-8578 Sendai, Japan 
\and Istituto ricerche solari Aldo e Cele Dacc\'o (IRSOL), Faculty of Informatics, Universit\`a della Svizzera italiana, Via Patocchi 57, Locarno, Switzerland}
 
\abstract
{We present the optical $BVR$ polarization measurements for the Be--X-ray pulsar binary \lsv from its giant outburst in early 2023 and subsequently over three years. We detect a significant polarization of about $\approx$1.1--2.1\%, which includes the interstellar contribution. Polarization evolves significantly with time since the outburst. The interstellar polarization was found to be $\approx$2.5--2.9\%. The obtained intrinsic optical polarization of \lsv after correction is between 0.5--1.7\%, perpendicular to the interstellar polarization direction. We consider various plausible sources of optical polarization and infer scattering in the Be decretion disk to be the most dominant source based on time and spectral evolution. Changing intrinsic PA with time hints at a plausible precession of the Be disk.}

   \keywords{Be/X-ray binaries -- pulsars: individual: LS V +44 17, RX J0440.9+4431 -- polarization -- stars: neutron 
               }

   \maketitle
%
%-------------------------------------------------------------------
\nolinenumbers

\section{Introduction}

X-ray binaries are binary star systems in which a compact object accretes matter from a stellar companion, typically a main sequence donor star.
These systems provide valuable insights into accretion processes that occur in the vicinity of compact objects where matter is subject to extreme magnetic and gravitational fields. 
In case of Be--X-ray pulsar binaries \citep[see][for reviews]{Reig2011review, reig2026review}, the donor is a Be star and the accretor is a neutron star (NS) operating as an X-ray pulsar \citep[XRP, see][for review]{MushtukovTsygankov2024}. 
The Be star is surrounded by a geometrically thick quasi-Keplerian decretion disk \citep{Lee1991}. 
A close encounter of the NS with the Be disk during the course of its orbit results in the so-called Type I outbursts with X-ray luminosities $\approx$10$^{36}$--10$^{37}$~erg~s$^{-1}$ due to increased accretion \citep{Okazaki2013}. 
Typically, these outbursts occur once per orbit.
Brighter type II or giant outbursts with X-ray luminosity exceeding $10^{37}$~erg~s$^{-1}$ are also possible in Be XRPs and are associated with the warping of Be disks \citep{Stella1986,Negueruela1998,Okazaki2013}. 
Generally, such outbursts last longer than Type I outbursts and occur once every few years.

\lsv (RX J0440.9$+$4431) is a Be--XRP binary \citep{Motch1997} situated at a distance, $d \approx 2.4$~kpc \citep{BailerJones2021}. 
\citet{reig1999discovery} confirmed the source as an XRP with a pulsar period of $202.5 \pm 0.5$~s. 
\citet{Reig2005} have extensively studied the companion star in the system. They identified the donor as a Be star of type B0.2V. 
The consistent presence of the H$\alpha$ line in the optical spectra, as well as its correlation with infrared magnitudes led to the conclusion that there exists long term optical and infrared variability in the system. 
Both observational signatures, the emission lines and the infrared excess, are thought to be due to the presence of an equatorial decretion disk, plausibly produced as a result of the rapid rotation of the star \citep{PorterRivinius2003}. The variability suggests long term structural changes in the Be disk.
\lsv went into type I outburst in 2010 and 2011, during which the orbital period was measured at $\approx$155~d  \citep{Morii2010,Finger2010} and the magnetic field was estimated at $\approx 3 \times 10^{12}$~G \citep{Tsygankov2012} using timing and spectral X-ray analyzes. The orbital period was later updated to $150.0\pm0.2$~days by \citet{Ferrigno2013}.

More recently, the source went into a similar outburst in December 2022 which was followed by a giant outburst in early 2023 \citep{Nakajima2022,Pal2023,Salganik2023ATel}. 
During this outburst, the source was also observed with the \textit{Imaging X-ray Polarimetry Explorer} \citep{Weisskopf2022} and X-ray linear polarization in the 2--8~keV band was measured by  \citet{Doroshenko2023}. 
Polarimetry is a unique technique to probe the geometry of the emitting region. 
XRPs are expected to have a high degree of polarization (PD) in X-rays due to the strong birefringence of magnetized plasma in the presence of strong magnetic fields \citep{Gnedin1978}. 
The electric vector position angle (aka polarization angle, PA) is thought to be related to the projection of the magnetic dipole of the XRP on the sky plane and is hence expected to evolve with the pulsar phase, as described by the rotating vector model \citep[RVM;][]{RC69,Meszaros88,poutanen20}. 
The geometry of the X-ray source in \lsv was interpreted using a two-component model \citep{Doroshenko2023,Zhao2025}. The first component that depends on the pulsar period was described using the RVM. In addition, a second component, constant with respect to the pulsar phase, is  potentially produced due to the scattering of the pulsar radiation in the accretion disk wind \citep{Nitindala2025}. Its presence was required to describe the evolution of X-ray polarization between supercritical and subcritical accretion states.

In this paper, we focus on the optical linear polarization of \lsv in the $BVR$ bands. 
The polarized optical emission in such systems can originate from various components such as the Be star and its decretion disk, the NS accretion disk and disk wind, reprocessed X-ray emission from the XRP in the disk and/or the companion star, and from the strong magnetic field of the NS and the interaction of the Be disk material with this magnetic field. Some of these may in turn show variability on different timescales. The aim of this work is to examine the source of optical polarization in \lsv and discuss its implications on the geometry of the binary system. 
The paper is organized as follows. 
The data is described in Sect.~\ref{sec:data}. 
The results are presented in Sect.~\ref{sec:results} and are discussed in detail in Sect.~\ref{sec:discussion}. 
Lastly, we summarize our findings in Sect.~\ref{sec:summary}.

\section{Data}\label{sec:data}

Optical polarimetric data of \lsv were obtained using two copies of DIPol-2 polarimeters \citep{Piirola2014Dipol2} mounted on the Tohoku 60 cm (T60) telescope at Haleakala Observatory, Hawaii and the 1-m telescope (C2PU) at Calern Observatory, Nice and using DIPol-UF polarimeter mounted on the Nordic Optical Telescope (NOT), La Palma. 
The DIPol polarimeters achieve high precision by splitting the incident optical light into two orthogonally polarized beams (e- and o-rays) which are then recorded as two points on a single CCD image. 
Simultaneous recording of two perpendicularly polarized beams removes, at an instrumental level, the errors induced due to atmospheric transparency and seeing variations. 
The effect of sky polarization is canceled out because of the equal intensities of the overlapping orthogonally polarized sky images such that the final stellar polarization obtained is free of any systematic error arising due to the sky polarization.
Thus, the accuracy of DIPol instruments is only photon-limited and for sufficiently bright sources, it can reach up to $10^{-5}$.
Such images are simultaneously observed in three bands ($B$, $V$, and $R$) with the use of two dichroic sharp-pass filters.

\lsv was observed for 17 nights between February 2023 and December 2025 (see Table~\ref{table:obs}). 
Depending on the sky conditions, 8--68 measurements of the target normalized Stokes parameters ($q$ and $u$) per night were taken simultaneously in $BVR$ bands with the exposure times of 5--30~s. 
From their nightly means \citep[computed using a $2\sigma$ weighting algorithm as described in][]{Kosenkov2017,Piirola2020}, the average PD 
\begin{equation} 
\label{eq:PD}
P=\sqrt{q^2 + u^2} 
,\end{equation}
and PA 
\begin{equation} 
\label{eq:PA}
\theta=\frac{1}{2}{\rm atan2}(u,q) , 
\end{equation}
were computed, where atan2 is the quadrant-preserving arctangent function. 
The uncertainty of the Stokes parameters is a weighted standard deviation of their average values and is the same as the error on the PD ($\sigma_P$). The error on PA is given as $\sigma_{\theta} = \sigma_P/(2P)$ \citep{Clarke1986,Naghizadeh1993}. 

\begin{figure*}%[ht!]
   \sidecaption %   \centering
   \includegraphics[width=12.5cm]{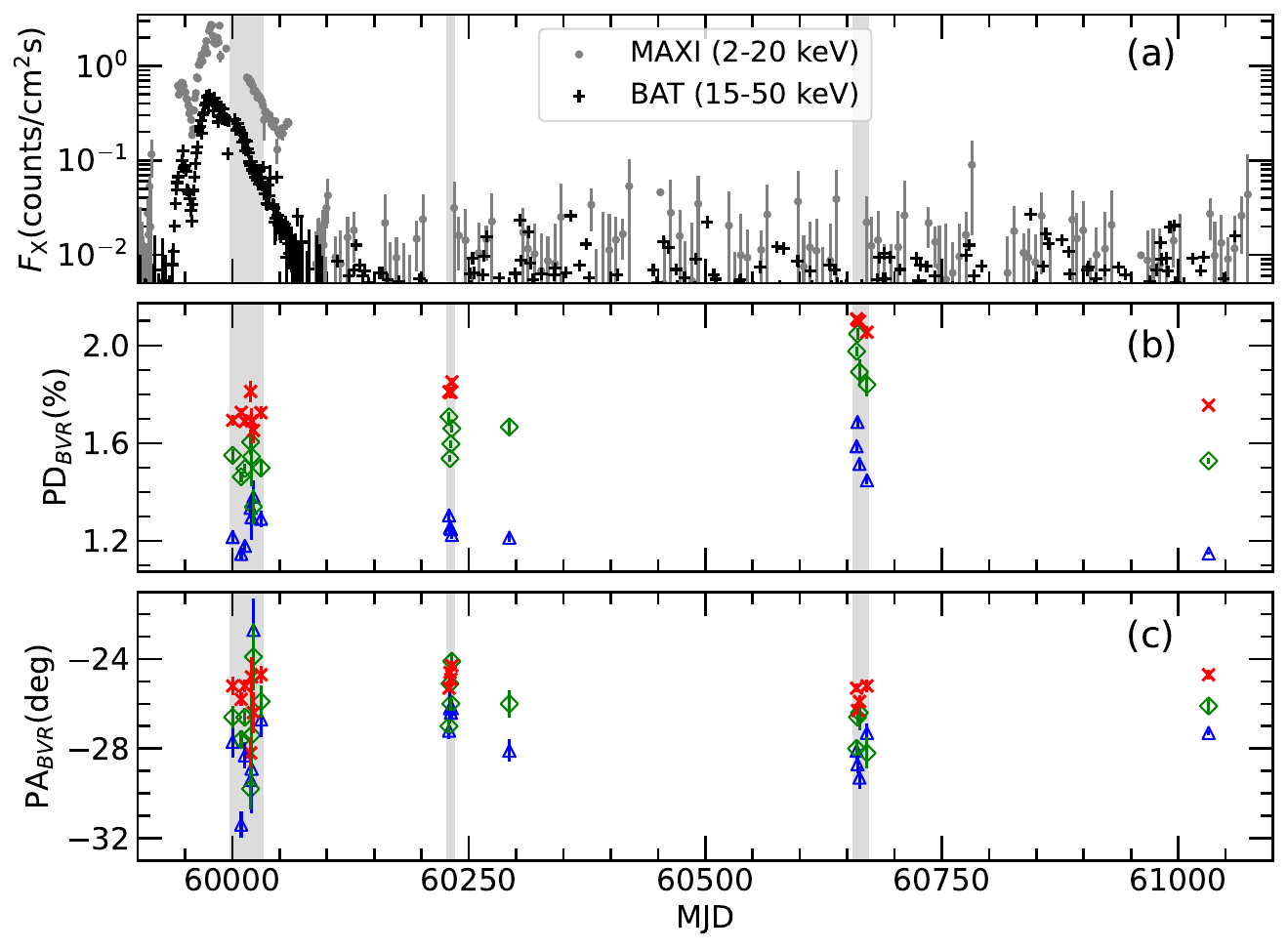}
      \caption{(a) MAXI and Swift/BAT X-ray light curves. Data outside the outburst are binned at seven and one-day intervals, respectively. Evolution of the observed (b) PD and (c) PA of \lsv with time in different filters: $B$ (blue triangles), $V$ (green diamonds), and $R$ (red crosses). The gray regions from left to right denote epochs 1, 2 and 3, respectively. 
              }
         \label{fig:timeEvol}
  \end{figure*}

The instrumental polarization for each telescope setup was determined by observing standard non-polarized stars listed in \citet{Piirola2020}, and was found to be at the level of 0.001\%, i.e. negligible for our purposes.
The zero point of the PA was determined by observing highly polarized stars (with T60 and C2PU, HD 204827 and HD 25443; with NOT, HD~204827 and BD~+59~389). 
Detailed descriptions of observation and calibration procedures are given in \citet{Kosenkov2017,Piirola2020}. 
The nightly averaged observed values of PD and PA are shown in Fig.~\ref{fig:timeEvol} and given in Table~\ref{table:obs}. 
In addition to the target, a few field stars were also observed (see Table~\ref{table:ism}) to determine the interstellar polarization. 
This is discussed in detail in Sect.~\ref{subsec:ISM}.

% Figure 1: Observed data
 %  \begin{SCfigure*}
%   \end{SCfigure*}

% Table 1: Observed data
\begin{table*}
\caption{Observed polarimetric data for \lsv.}             
\label{table:obs}      
\centering          
\begin{tabular}{l c c c c c c c}     % 7 columns 
\hline\hline       
                      % To combine 4 columns into a single one 
  & \multicolumn{2}{c}{$B$} & \multicolumn{2}{c}{$V$} & \multicolumn{2}{c}{$R$} &\\ 
%\hline
MJD & PD (\%) & PA (deg) & PD (\%) & PA (deg) & PD (\%) & PA (deg) & Telescope \\

\hline
   60000.3594 & 1.22 $\pm$ 0.03 & $-$27.7 $\pm$ 0.7 & 1.55 $\pm$ 0.03 & $-$26.6 $\pm$ 0.5 & 1.69 $\pm$ 0.02 & $-$25.2 $\pm$ 0.4 & T60 \\ 
   60009.3725 & 1.15 $\pm$ 0.02 & $-$31.4 $\pm$ 0.6 & 1.46 $\pm$ 0.02 & $-$27.6 $\pm$ 0.4 & 1.73 $\pm$ 0.02 & $-$25.8 $\pm$ 0.3 & \\
   60013.2897 & 1.18 $\pm$ 0.02 & $-$28.3 $\pm$ 0.6 & 1.50 $\pm$ 0.02 & $-$26.6 $\pm$ 0.4 & 1.69 $\pm$ 0.02 & $-$25.2 $\pm$ 0.3 & \\ 
   60019.3568 & 1.33 $\pm$ 0.04 & $-$29.4 $\pm$ 0.9 & 1.60 $\pm$ 0.05 & $-$29.8 $\pm$ 0.9 & 1.81 $\pm$ 0.05 & $-$28.2 $\pm$ 0.7 & \\ 
   60020.3403 & 1.30 $\pm$ 0.09 & $-$28.9 $\pm$ 2.0 & 1.54 $\pm$ 0.12 & $-$27.4 $\pm$ 2.3 & 1.69 $\pm$ 0.05 & $-$24.8 $\pm$ 0.9 & \\ 
   60022.3592 & 1.38 $\pm$ 0.07 & $-$22.7 $\pm$ 1.4 & 1.34 $\pm$ 0.07 & $-$23.9 $\pm$ 1.5 & 1.65 $\pm$ 0.05 & $-$26.4 $\pm$ 0.9 & \\ 
   60030.3193 & 1.29 $\pm$ 0.03 & $-$26.7 $\pm$ 0.8 & 1.50 $\pm$ 0.04 & $-$25.9 $\pm$ 0.7 & 1.72 $\pm$ 0.03 & $-$24.7 $\pm$ 0.4 & \\

\hline
 Epoch 1 & 1.210 $\pm$ 0.001 & $-$28.70 $\pm$ 0.02 & 1.496 $\pm$ 0.001 & $-$26.97 $\pm$ 0.02 & 1.710 $\pm$ 0.001 & $-$25.49 $\pm$ 0.02 & \\ %-- for first set of observations only
   
\hline

   60229.0619 & 1.30 $\pm$ 0.02 & $-$27.2 $\pm$ 0.4 & 1.71 $\pm$ 0.02 & $-$27.0 $\pm$ 0.3 & 1.81 $\pm$ 0.01 & $-$25.3 $\pm$ 0.2 & C2PU \\
   60230.1053 & 1.25 $\pm$ 0.03 & $-$26.2 $\pm$ 0.7 & 1.54 $\pm$ 0.02 & $-$25.1 $\pm$ 0.3 & 1.81 $\pm$ 0.02 & $-$24.6 $\pm$ 0.3 & \\
   60231.0805 & 1.25 $\pm$ 0.01 & $-$26.4 $\pm$ 0.3 & 1.60 $\pm$ 0.02 & $-$26.0 $\pm$ 0.3 & 1.81 $\pm$ 0.02 & $-$24.9 $\pm$ 0.4 & \\
   60232.0644 & 1.22 $\pm$ 0.02 & $-$26.2 $\pm$ 0.4 & 1.66 $\pm$ 0.02 & $-$24.1 $\pm$ 0.3 & 1.85 $\pm$ 0.02 & $-$24.3 $\pm$ 0.2 & \\

\hline

Epoch 2  & 1.253 $\pm$ 0.001 & $-$26.50 $\pm$ 0.02 & 1.615 $\pm$ 0.001 & $-$25.51 $\pm$ 0.02 & 1.822 $\pm$ 0.001 & $-$24.80 $\pm$ 0.02 & \\
\hline       
60292.8411 & 1.21 $\pm$ 0.02 & $-$28.1 $\pm$ 0.5 & 1.67 $\pm$ 0.03 & $-$26.0 $\pm$ 0.6 &  --  &  --  & \\
\hline

60659.9331 & 1.59 $\pm$ 0.02 & $-$28.1 $\pm$ 0.4 & 1.98 $\pm$ 0.02 & $-$28.0 $\pm$ 0.3 & 2.11 $\pm$ 0.01 & $-$25.3 $\pm$ 0.2 & \\
    60660.9030 & 1.69 $\pm$ 0.02 & $-$28.7 $\pm$ 0.4 & 2.05 $\pm$ 0.02 & $-$26.6 $\pm$ 0.3 & 2.10 $\pm$ 0.02 & $-$26.3 $\pm$ 0.2 & \\
60662.9147 & 1.51 $\pm$ 0.02 & $-$29.3 $\pm$ 0.5 & 1.89 $\pm$ 0.05 & $-$26.4 $\pm$ 0.8 & 2.10 $\pm$ 0.02 & $-$25.9 $\pm$ 0.3 & \\
    60670.8992 & 1.45 $\pm$ 0.02 & $-$27.3 $\pm$ 0.4 & 1.84 $\pm$ 0.05 & $-$28.2 $\pm$ 0.7 & 2.05 $\pm$ 0.02 & $-$25.2 $\pm$ 0.3 & \\

\hline
Epoch 3  & 1.536 $\pm$ 0.001 & $-$28.21 $\pm$ 0.02 & 1.994 $\pm$ 0.001 & $-$26.78 $\pm$ 0.02 & 2.093 $\pm$ 0.001 & $-$25.98 $\pm$ 0.01 & \\
\hline  

\hline
%NOT Dec 2025 
61032.0405 & 1.148 $\pm$ 0.006 & $-$27.3 $\pm$ 0.1 & 1.53 $\pm$ 0.01 & $-$26.1 $\pm$ 0.3 & 1.76 $\pm$ 0.01 & $-$24.7 $\pm$ 0.2 & NOT \\
\hline

\end{tabular}
\end{table*}

\section{Results}\label{sec:results}

\subsection{Observed polarization}\label{subsec:obs_pol}
The observed polarization of \lsv is shown in Fig.~\ref{fig:timeEvol}b,c. The first set of observations (Epoch 1) between MJD~60000 and 60030 was taken during the decay of the Type II outburst that occurred in early 2023. Subsequently, more observations were taken at times much after the giant outburst and we have grouped part of the data for further analysis. Epoch 2 consists of the data collected between MJD 60229 and 60232 while Epoch 3 consists of data obtained between MJD 60659 and 60670. In Fig.~\ref{fig:timeEvol}, we see that the observed PD shows an increase from Epoch 1 through Epoch 2 to Epoch 3. The observed data values and the weighted averages during each epoch are presented in Table~\ref{table:obs}. The observed PD seems to return to values similar to those during Epoch 1 towards the end of our observing campaign (MJD 61032).
Correspondingly, the PA fluctuates marginally between epochs. 

In order to investigate the significance of change in observed polarization values between the different epochs, we used the multi-variate Hotelling's $T^2$ test \citep{Hotelling1931}. The details of using this test on optical polarimetric data have been described in detail in \citet{Kosenkov2017}. Table~\ref{table:t2test} summarizes the results of this test between different epochs. We see that there are statistically significant changes in the observed polarization between different epochs (except between Epochs 1 \& 2 in the $B$ band).

% Table 2: Statistical test
\begin{table*}
\caption{Results of the Hotelling's $T^2$ test.}             
\label{table:t2test}      
\centering          
\begin{tabular}{l c c c c c c c}    
\hline\hline 
Epoch & Band & $t^2$ & $f$ & $p$ & $N$ & Significance & Inference \\
\hline
1 \& 2 & $B$ & 4.2 & 2.1 & 0.1273 & 272 & $0.127$ & Equal polarization \\
1 \& 2 & $V$ & 29.8 & 14.9 & $< 10^{-6}$ & 272 & $7.59 \times 10^{-7}$ & Unequal polarization \\
1 \& 2 & $R$ & 23.3 & 11.6 & $< 10^{-4}$ & 272 & $1.45 \times 10^{-5}$ & Unequal polarization \\
\hline
1 \& 3 & $B$ & 142.8 & 71.0 & $< 10^{-21}$ & 188 & $1.31 \times 10^{-23}$ & Unequal polarization \\
1 \& 3 & $V$ & 247.2 & 122.9 & $< 10^{-31}$ & 188 & $1.08 \times 10^{-34}$ & Unequal polarization \\
1 \& 3 & $R$ & 269.0 & 133.8 & $< 10^{-35}$ & 188 & $1.15 \times 10^{-36}$ & Unequal polarization \\
\hline
2 \& 3 & $B$ & 237.7 & 118.2 & $< 10^{-25}$ & 172 & $7.86 \times 10^{-33}$ & Unequal polarization \\
2 \& 3 & $V$ & 250.3 & 124.4 & $< 10^{-24}$ & 172 & $6.05 \times 10^{-34}$ & Unequal polarization \\
2 \& 3 & $R$ & 333.7 & 165.9 & $< 10^{-34}$ & 172 & $1.38 \times 10^{-40}$ & Unequal polarization \\
\hline
\end{tabular}
\tablefoot{$f$ is the $F$-statistic, $p$ is the one-sided p-value, and $N$ is the total number of observations including both sets of data. The significance denotes the probability that the two sets of observations have equal polarization.}
\end{table*}

\begin{figure}
   \centering
   \includegraphics[width=8.5cm]{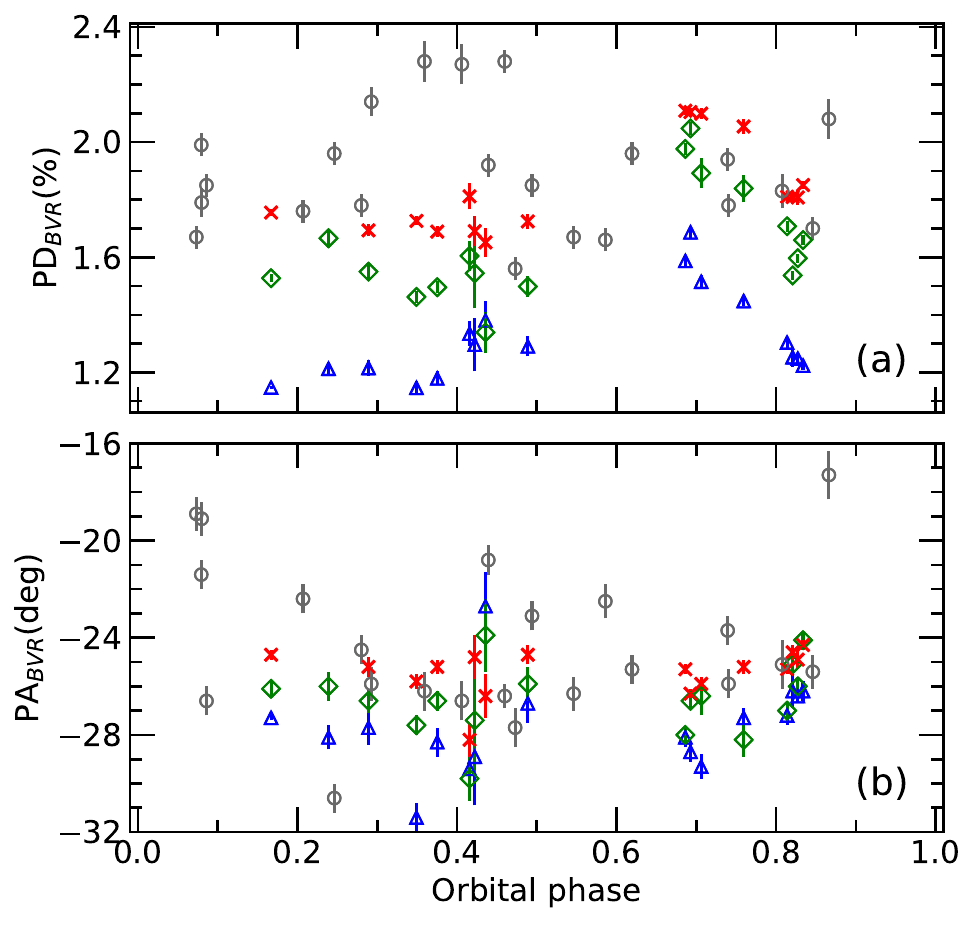}
      \caption{Observed (a) PD and (b) PA of \lsv as a function of the binary orbital phase in different filters: $B$ (blue triangles), $V$ (green diamonds), and $R$ (red crosses). RoboPol $R$ band measurements reported in \citet{Reig2026} are shown as gray circles.
              }
         \label{fig:PDPA_orbit}
   \end{figure}

To assess if this change is related to the orbital period \citep[$\approx 150$ days;][]{Ferrigno2013} of the binary, we plot the data against the orbital phase (Fig.~\ref{fig:PDPA_orbit}). Since the data we obtained using DIPol instruments do not cover all the orbital phases, we also overplot data presented in \citet{Reig2026} obtained using the RoboPol polarimeter (available only in $R$ band). We do not see any clear orbital phase dependence of polarization. 

% Figure 2: Orbital phase dependence 

\begin{figure}
   \centering
   \includegraphics[width=8.5cm]{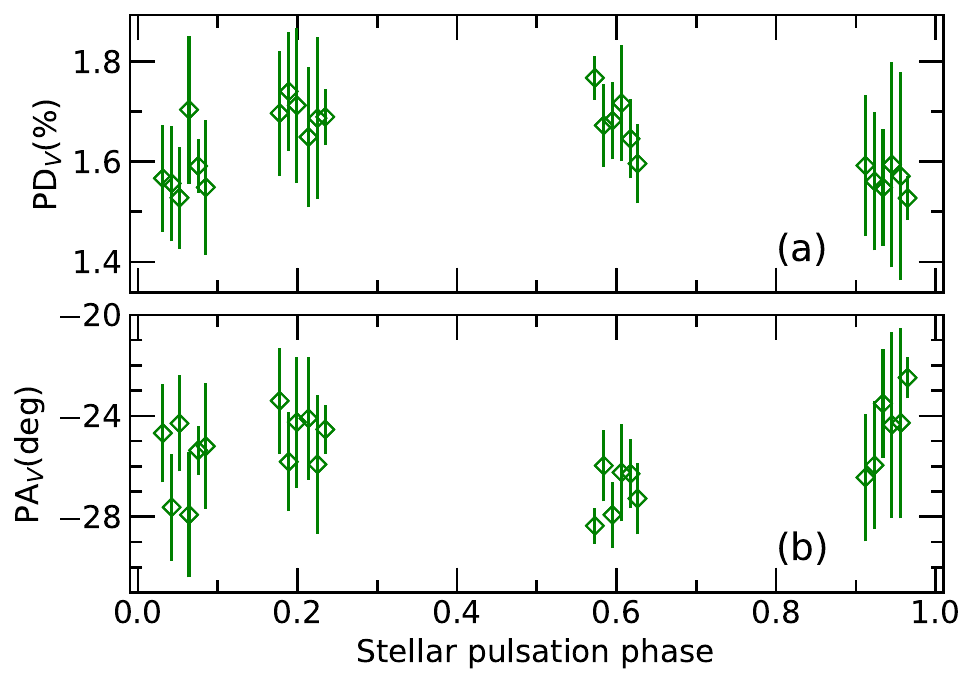}
      \caption{Observed (a) PD and (b) PA of \lsv during Epoch 2 as a function of the Be stellar pulsation phase in the $V$ band.
              }
         \label{fig:PDPA_stellar}
   \end{figure}

Next, we checked if there is any correlation of optical polarization to the stellar pulsation period \citep[$\approx 7.5$h;][]{reig2022fast} of the Be companion. Since the variability timescale is of the order of hours, nightly means cannot be folded to get the stellar phase dependence. Instead, we used individual measurements during each observing night. Moreover, we also separate data from different epochs to ensure that the larger timescale variation is excluded. Figure~\ref{fig:PDPA_stellar} shows the observed polarization measurements in the $V$ band taken during Epoch 2 grouped within a bin size of 0.01 of the stellar pulsation phase. Due to the sparse coverage of the data over the phase plot and the large error bars, any variability on this timescale remains inconclusive. 

% Figure 3: Stellar pulsation phase dependence 

\begin{figure}
\centering
\includegraphics[width=8.5cm]{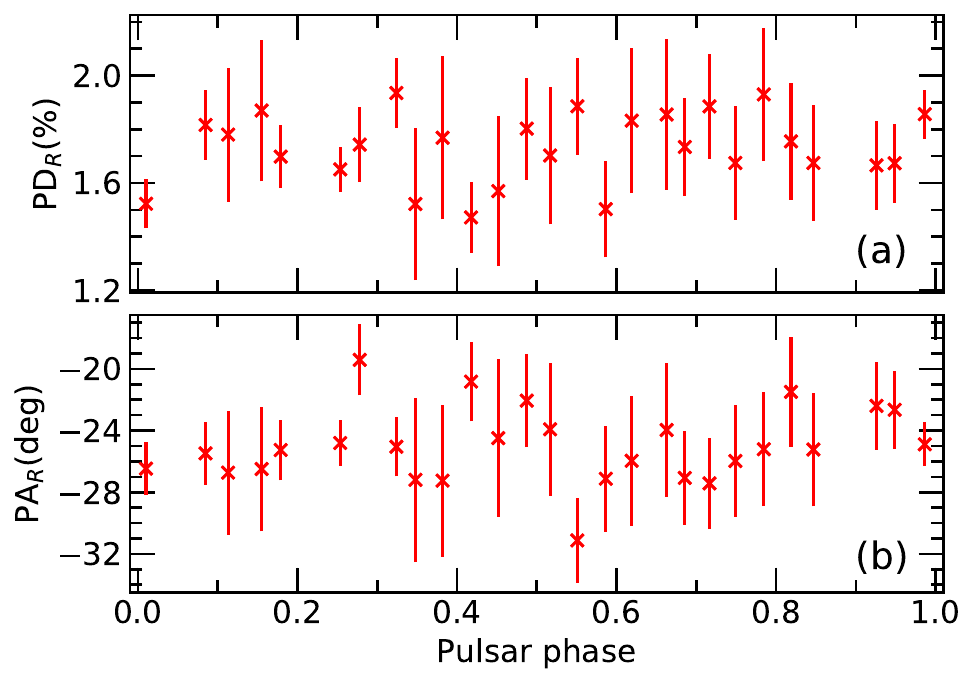}
\caption{Observed (a) PD and (b) PA of \lsv on MJD 60030 in the $R$ band as a function of the XRP spin phase.}
\label{fig:PDPA_pulsar}
\end{figure}

Lastly, we examined whether the optical polarization in the source was varying with the X-ray pulsar spin phase \citep[$\approx$204~s;][]{Doroshenko2023}. Again, we used individual measurements within an observing night in order to not smear out any possible variability. Furthermore, we only used observations with exposure time of 5~s (which was viable only for the data obtained on MJD~60019 and 60030). One set of measurements of the Stokes parameters was obtained in 20~s, i.e. $\approx 10\%$ of the spin period. Observations with larger exposure times cannot resolve variability on timescales of few hundreds of seconds. Figure~\ref{fig:PDPA_pulsar} shows individual polarization measurements in the $R$ band taken on MJD 60030 grouped within a phase bin size of 0.025. The PD and PA are consistent with a constant value throughout spin period. 
Thus, we conclude that the change in polarization between epochs is not dependent on orbital, stellar pulsation, and XRP spin timescales associated with the Be/X-ray pulsar binary system but rather simply a variation in time over a longer duration since the giant outburst.

% Figure 4: Pulsar phase dependence 

\subsection{Interstellar polarization}\label{subsec:ISM} 

\begin{figure}
   \centering
   \includegraphics[width=9cm]{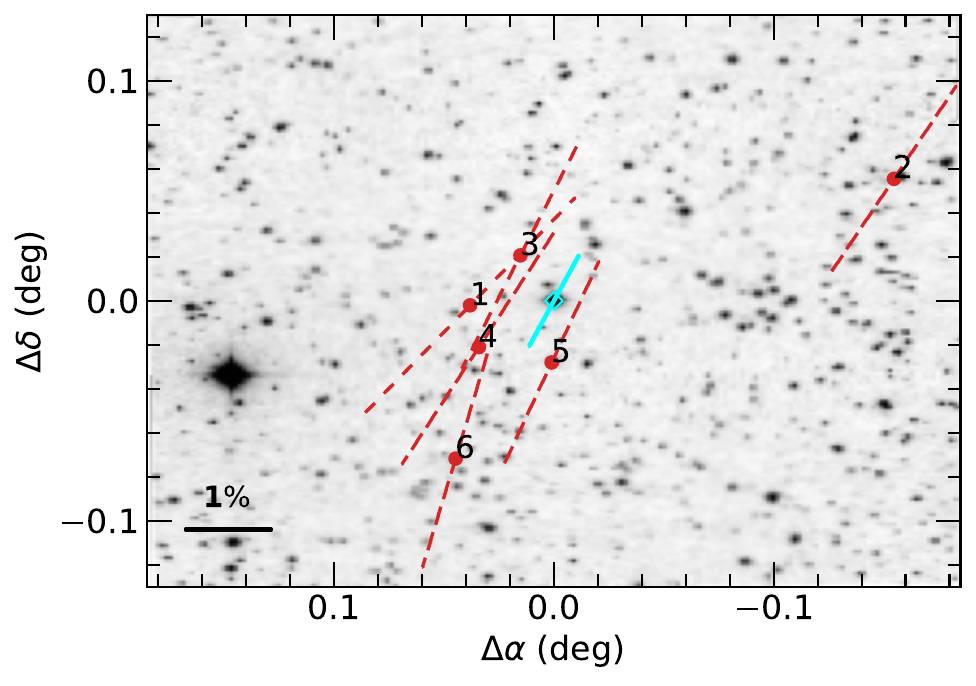}
      \caption{Polarization vectors of the field stars (red circles and dashed lines) and \lsv (cyan diamond) during Epoch 1 in the $B$ filter. The length of the bars corresponds to their PD and the direction of the bars corresponds to their PA. The coordinates are centered at \lsv ($\alpha = 70\fdg247$, $\delta = 44\fdg530$).}
         \label{fig:RAdec_B}
   \end{figure}

The observed measurements include not only polarization intrinsic to \lsv, but also a contribution from the interstellar (IS) medium due to dichroic absorption which must be accounted for. For this purpose, we studied the polarization of field stars within an angular distance of $16'$ around the target and at comparable parallaxes. This method of estimating the IS polarization has been shown to give robust results \citep{Kosenkov2017}. The resulting PD and PA of the field stars observations are given in Table~\ref{table:ism}. In Fig.~\ref{fig:RAdec_B}, we show a sky-map with the polarization vectors of the target and the field stars in the $B$ band. 

% Figure 5: Sky map

% Table 3: Field stars data
\begin{table*}
\caption{Polarimetric data for the field stars and their weighted average.}             
\label{table:ism}      
\centering          
\begin{tabular}{c c c c c c c c }     % 7 columns 
\hline\hline       
                      % To combine 4 columns into a single one 
  & & \multicolumn{2}{c}{$B$} & \multicolumn{2}{c}{$V$} & \multicolumn{2}{c}{$R$}\\ 
%\hline
Star & $\pi$ (mas) & PD (\%) & PA (deg) & PD (\%) & PA (deg) & PD (\%) & PA (deg) \\
\hline
1  & 0.22 $\pm$ 0.03 & 3.5 $\pm$ 1.7 & $-$44 $\pm$ 13 & 3.52 $\pm$ 0.30  & $-$37.9 $\pm$ 2.4 & 4.35 $\pm$ 0.17 & $-$38.2 $\pm$ 1.1 \\
    2  & 0.36 $\pm$ 0.02 & 2.60 $\pm$ 0.09  & $-$33.9 $\pm$ 1.0  & 2.61 $\pm$ 0.09  & $-$32.9 $\pm$ 1.0  & 2.62 $\pm$ 0.05 & $-$32.6 $\pm$ 0.6 \\
3  & 0.38 $\pm$ 0.02 & 3.00 $\pm$ 0.03  & $-$27.3 $\pm$ 0.3  & 2.47 $\pm$ 0.04 & $-$25.2 $\pm$ 0.4   & 2.655 $\pm$ 0.003 & $-$24.94 $\pm$ 0.04 \\

4  & 0.41 $\pm$ 0.02 & 3.27 $\pm$ 0.12  & $-$33.3 $\pm$ 1.0  & 2.74 $\pm$ 0.10 & $-32.8$ $\pm$ 1.1 & 2.76 $\pm$ 0.04  & $-$32.0 $\pm$ 0.5   \\
    5  & 0.52 $\pm$ 0.02  & 2.599 $\pm$ 0.005  & $-$25.05 $\pm$ 0.05  & 2.30 $\pm$ 0.09 & $-$27.9 $\pm$ 1.1 & 2.39 $\pm$ 0.10 & $-$23.6 $\pm$ 1.2 \\
6  & 0.55 $\pm$ 0.02 & 2.7 $\pm$ 0.6  & $-$16.9 $\pm$ 6.4  & 2.63 $\pm$ 0.09 & $-$25.6 $\pm$ 0.9 & 2.84 $\pm$ 0.04  & $-$21.4 $\pm$ 0.4 \\
\hline 

2--4 &  & 2.957 $\pm$ 0.004 & $-$28.04 $\pm$ 0.04 & 2.505 $\pm$ 0.007 & $-$27.2 $\pm$ 0.1 &  2.655 $\pm$ 0.001 & $-$25.02 $\pm$ 0.01 \\
\hline
\end{tabular}
\tablefoot{The parallax of \lsv is $\pi=0.38 \pm 0.02$ mas.}
\end{table*}

\begin{figure*}
   \centering
   \includegraphics[width=18cm]{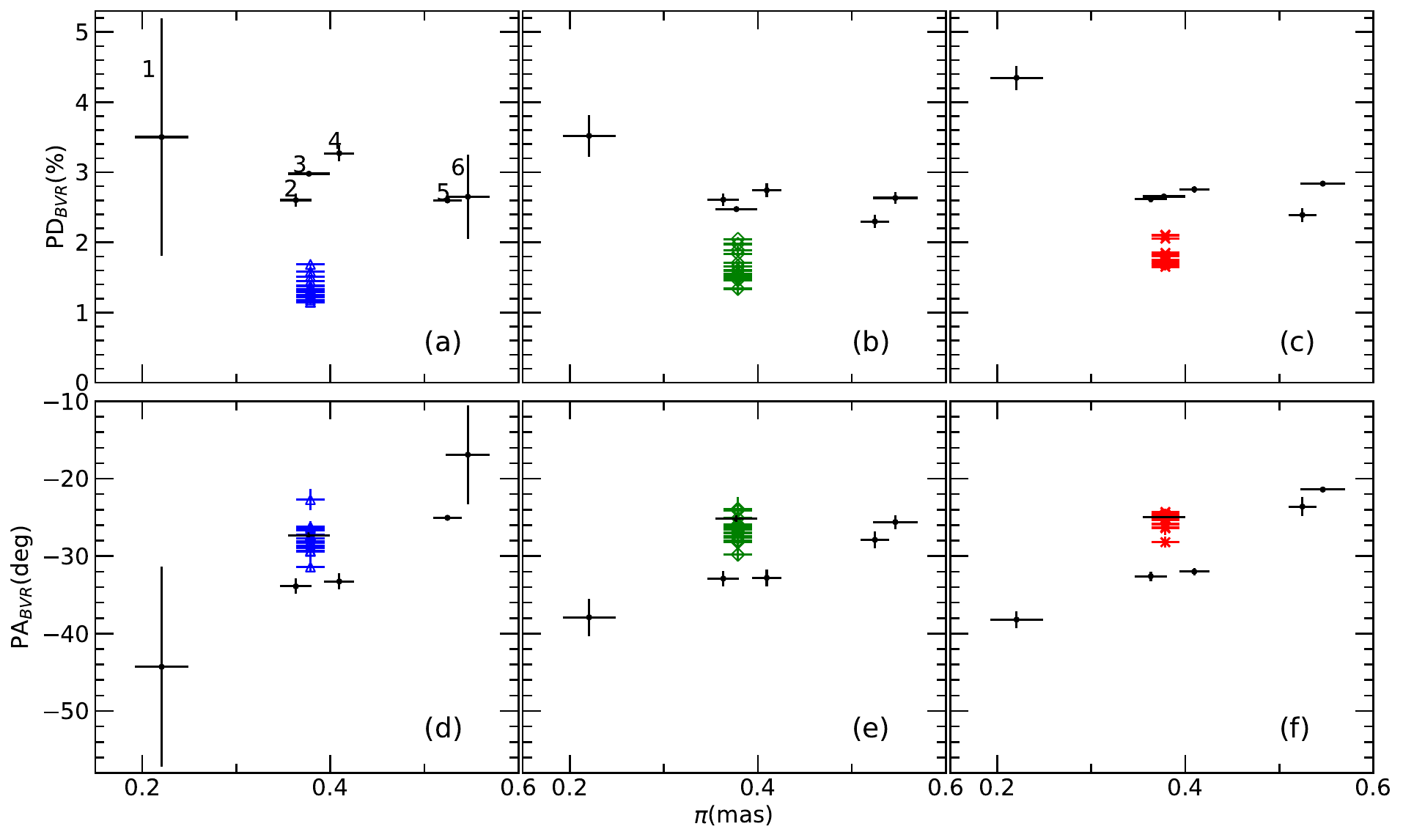}
      \caption{Observed (a--c) PD and (d--f) PA as a function of parallax ($\pi$) for \lsv (colored symbols) and the field stars (black circles) in the $B$ (left panels), $V$ (middle panels), and $R$ (right panels) bands.}
    \label{fig:PDPAparallax_BVR}
   \end{figure*}

In Fig.~\ref{fig:PDPAparallax_BVR}, the PD and PA for each of these stars is shown as a function of parallax (obtained from \textit{Gaia}\footnote{\url{https://www.cosmos.esa.int/web/gaia}} Data Release 3). The PD of the field stars is decreasing with increase in parallax in all three bands. The PD of the target, \lsv, differs from those of the field stars at a similar parallax. On the other hand, the PA of the target falls in line with the field stars PA dependence on parallax, which is increasing with parallax. For our estimate of the IS polarization, we choose the three field stars with the closest parallax to the target (that is, stars 2--4 from Table~\ref{table:ism}) and analyze them further.

% Figure 6: Field stars PD and PA with parallax

\begin{figure*}
   \centering
   \includegraphics[width=18cm]{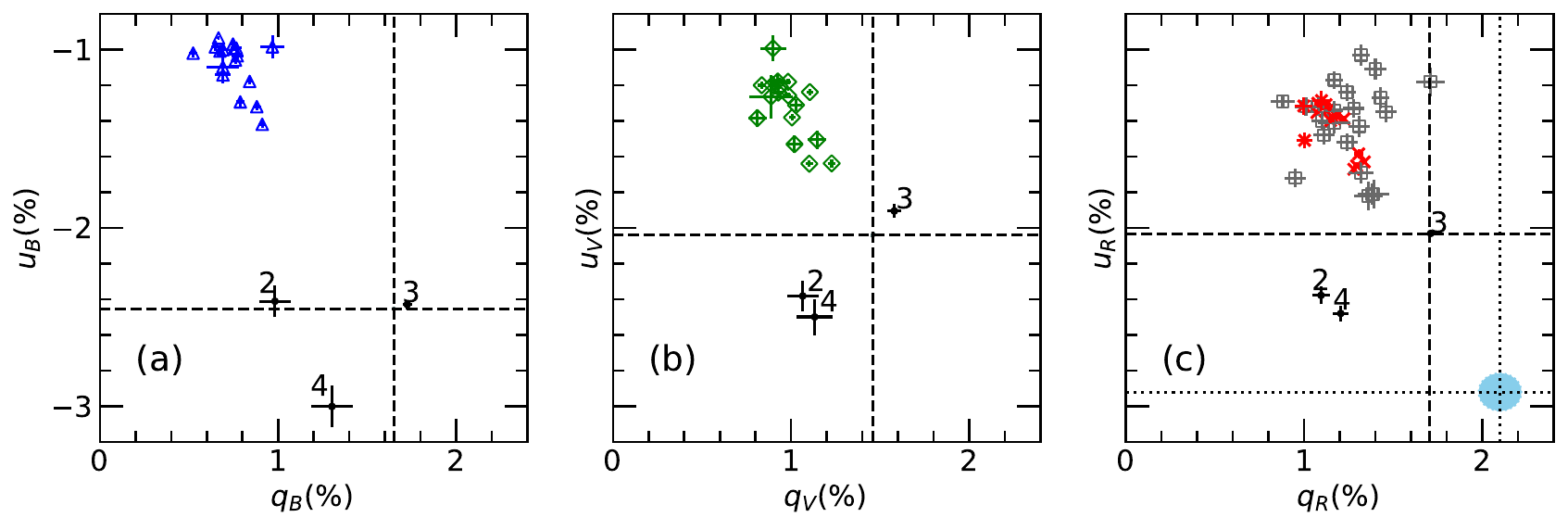}
\caption{Observed normalized Stokes $q$ and $u$ parameters of \lsv (colored symbols) and field stars 2--4 (black crosses) in (a) $B$, (b) $V$, and (c) $R$ bands. The average of the field stars are shown by the intersection of the dashed lines with the errors being too small to be perceptible. 
In panel (c), observations of \lsv taken using the RoboPol instrument are shown as gray squares \citep{Reig2026}. Their IS polarization estimate is shown as a blue circle at the intersection of dotted lines. The errors on all observed data (field stars included) are at $1\sigma$ level whereas the errors on all of the averaged IS polarization values are at $3\sigma$ level.
              }
         \label{fig:quPlane}
   \end{figure*}

The observed normalized Stokes parameters for \lsv and the three chosen field stars are shown in Fig.~\ref{fig:quPlane}. In all three filters, the Stokes parameters of the target cluster in a region separate from region occupied by the field stars on the $q$--$u$ plane, indicating that \lsv has a significantly distinguished intrinsic polarization. The weighted average polarization of the three field stars (as also given in Table~\ref{table:ism}) is shown by the intersection of the dashed lines. The PA of this estimate is along the direction of change of the observed PA of \lsv, thereby substantiating our considered value of IS polarization. Furthermore, in the $R$ band, we can compare our IS polarization estimate with that obtained in \citet{Reig2026} (shown by blue circle at the intersection of the dotted lines). We see that the PA is again in line with both, the direction of change of observed polarization of \lsv and our estimate. However, the IS PD is different from our estimate. This is likely due to the different choice of field stars. \citet{Reig2026} chose four field stars, all of which were slightly farther away from the target itself ($0.29 \lesssim \pi \lesssim 0.38$). The extinction around these distances seems to have a sharp increase with distance \citep[see figure B.1 in ][]{Reig2026} and hence IS polarization that is thought to be introduced due to selective absorption in the dust can also vary largely based on the distance to the field stars. 

% Figure 7: qu plane with target and close field stars

Star \#3 in Table~\ref{table:ism} is close to \lsv both in terms of the radial distance from the observer as well as the angular distance on the sky (Figs.~\ref{fig:RAdec_B} and \ref{fig:PDPAparallax_BVR}). Thus, we can regard it as a good representative field star for the IS estimate. Our weighted average IS polarization also lies close to this value. Choosing the IS estimate given in \citet{Reig2026} would only increase the intrinsic PD by $\leq1\%$ (which is also consistent with the RoboPol instrumental polarization at $3\sigma$ level) and maintain the same intrinsic PA. Hence, qualitative aspects of the resulting intrinsic polarization would not change by the choice of IS polarization. In this work, we consider the weighted average polarization of stars \#2--4 as the IS polarization value for subsequent discussion.

\subsection{Intrinsic polarization} 

The intrinsic polarization was computed by subtracting the Stokes parameters of the IS polarization from those of the observed polarization. The intrinsic PD and PA for the three epochs are given in Table~\ref{table:intrinsic}. The PD in the $B$ band is about twice as large as the PD in the $V$ and $R$ bands during all epochs. The PA is along the direction of change on the $q$--$u$ plane in all bands and is orthogonal to the IS PA. Since the IS polarization is considered to be a constant on such timescales, the change in observed polarization discussed in Sect.~\ref{subsec:obs_pol} can be taken to be solely due to the change in intrinsic polarization of \lsv. The intrinsic PD in each band decreases from Epoch 1 to Epoch 3, while the changes in the intrinsic PA are marginal, first decreasing during Epoch 2 and then increasing again during Epoch 3. 

The exact values of intrinsic PD and and PA depend considerably on the systematic uncertainties in the IS polarization. For example, if we use the IS polarization estimate from \citet{Reig2026}, the intrinsic PD in the $R$ band is greater by $\approx 0.95\%$ while the difference in PD between epochs remains the same. This is as expected since the choice of IS estimate only shifts the magnitude of Stokes parameters on the $q$--$u$ plane but not the difference between them. The intrinsic $R$-band PA decreases by $\approx 4\degr$--$7\degr$. Albeit a similar trend of decrease (Epoch 2) and subsequent increase (Epoch 3) is still seen in the intrinsic PA, the difference in PA values between epochs is largely reduced. Thus, while the choice of IS polarization is critical to deducing intrinsic polarization, we infer that any revised intrinsic PD and PA values do not qualitatively change our interpretation of the same. 

\begin{figure}
   \centering
   \includegraphics[width=9cm]{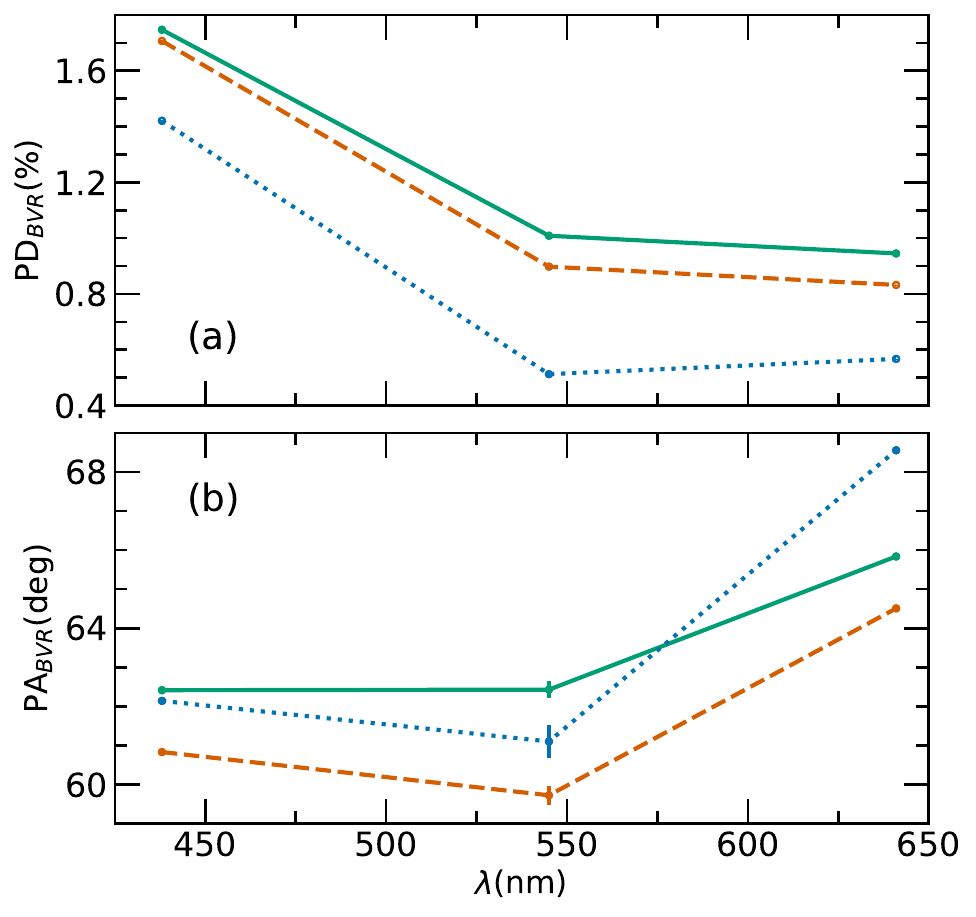}
\caption{Intrinsic (a) PD and (b) PA of \lsv as a function of wavelength during the three epochs. Epoch 1 is denoted by the solid, green line; Epoch 2 by the dashed, orange line; and Epoch 3 by the dotted, blue line.}
\label{fig:PDPA_wavelength}
\end{figure}

The wavelength dependence of intrinsic PD and PA during the three epochs is shown in Fig.~\ref{fig:PDPA_wavelength}. In all cases, there is a decreasing trend of PD from the $B$ band towards the $R$ band. The PA increases from the $B$ band towards the $R$ band. The increase in $R$ band PA during Epoch 3 is sharper than during other epochs while the spectral dependence of PD is similar during all epochs.

% Figure 8: Wavelength dependence of intrinsic PD and PA

\begin{table*}
\caption{Intrinsic polarization of \lsv.}             
\label{table:intrinsic}      
\centering          
\begin{tabular}{l c c c c c c }     % 7 columns 
\hline\hline       
                      % To combine 4 columns into a single one 
  & \multicolumn{2}{c}{$B$} & \multicolumn{2}{c}{$V$} & \multicolumn{2}{c}{$R$}\\ 
%\hline
Epoch & PD (\%) & PA (deg) & PD (\%) & PA (deg) & PD (\%) & PA (deg) \\
\hline

Epoch 1 & 1.747 $\pm$ 0.004 & 62.41 $\pm$ 0.07 & 1.009 $\pm$ 0.007 & 62.4 $\pm$ 0.2 & 0.945 $\pm$ 0.001 & 65.84 $\pm$ 0.04 \\

Epoch 2 & 1.707 $\pm$ 0.004 & 60.83 $\pm$ 0.07 & 0.898 $\pm$ 0.007 & 59.7 $\pm$ 0.2 & 0.832 $\pm$ 0.001 & 64.51 $\pm$ 0.05 \\

Epoch 3 & 1.421 $\pm$ 0.004 & 62.14 $\pm$ 0.09 & 0.513 $\pm$ 0.007 & 61.1 $\pm$ 0.4 & 0.567 $\pm$ 0.001 & 68.56 $\pm$ 0.07 \\
\hline                  
\end{tabular}
\end{table*}

\section{Discussion}\label{sec:discussion}
\subsection{Principal sources of polarized optical emission in Be--XRP binaries}

The spectral dependence of the intrinsic PD and PA, along with their evolution with time, can be used to constrain the origin of optical emission from Be--XRP binaries.

\subsubsection{Emission from the Be companion}
Optical emission from Be/X-ray binaries is dominated by the Be star and decretion disk system. In general, the optical starlight from the Be star can itself be polarized at a sub-percent level if it undergoes non-radial stellar pulsations or polarization can be observed due to electron scattering of the stellar light off the circumstellar envelope seen in Be stars \citep[that is, either the equatorial decretion disk or the polar stellar winds; ][]{Nagirner62, LoskutovSobolev1979, LoskutovSobolev1981, quirrenbach1997constraints, reig2018warped}.  
The PD produced by single electron scattering in an axially symmetric circumstellar envelope is of the order of a few percent and depends on the prevalent opacity, oblateness of the envelope, and the inclination \citep{brown1977polarisation}. The PA is perpendicular to the major axis of the oblate or prolate envelope. 
The Be stellar light can undergo multiple scatterings off the decretion disk, also producing polarization, with PD $< 4\%$ and PA along the decretion disk normal \citep{LS85}.
In the wavelength range relevant to our observations (wavelengths redder than the Balmer jump at 364.5~nm), PD decreases with increase in wavelength. However, since the PA depends only on the geometry of the envelope, it is expected to be independent of the wavelength. 

Furthermore, it has been observed that Be stars, either individual or in binaries, have highly variable optical polarization on timescales of years \citep{quirrenbach1997constraints, reig2018warped, Reig2026}. In the case of an isolated Be star Pleione, variable polarization \citep{hirata2007disk} was seen to be a result of the precession of the Be decretion disk, which is also common in Be stars residing in binary systems \citep{hummel1998spectacular,Martin2011,Martin2023}. Long-term polarimetric variability study of the Be--X-ray binary, X-Per \citep{kunjaya1995multi, roche1997observations} showed that the variability timescales are in agreement with what is expected from a decretion disk undergoing its formation (optical high state) and dissipation (optical low state) phases. Be binaries can also show polarimetric variability periodic at orbital timescales \citep[$\approx$ months][]{kravtsov2020orbital}. Short term variations seen as stellar pulsations originating from the interior of the Be star have also been observed \citep{reig2022fast} and this can also translate to variability in polarization on similar timescales ($\approx$ hours).

\subsubsection{Emission from the NS}
Additionally, the XRP may also contribute to optical emission and polarization. When the NS passes through the periastron of the binary orbit, the accretion process can be triggered. This results in enhanced X-ray emission from the NS and its accretion disk. Alternatively, warping of the Be disk towards the plane of NS orbital motion can also trigger accretion where the Be decretion disk acts as a reservoir of matter that can be accreted onto the XRP. The resulting accretion disk can produce optical emission from the outer radii. This emission can be polarized up to $\approx12\%$ along the plane of the optically thick disk depending on the inclination \citep{Cha60,Sob63} and is expected to have a blue spectrum.
In addition, X-ray emission can also be reprocessed in the outer regions of the accretion disk or in the Be decretion disk, resulting in polarized optical emission \citep{kunjaya1995multi, roche1997observations, reig2018warped}. In this case, optical polarization may follow the variability of the irradiating XRP emission (on the timescale of $\approx$ seconds) depending on the distance of the optically emitting region from the XRP. An irradiated disk is expected to have a flatter spectrum.

Lastly, any matter traversing the magnetic field lines in the vicinity of the XRP might produce synchrotron radiation that is inherently polarized perpendicular to the magnetic field lines. In that case, the PD depends on the ordering of the magnetic field, theoretically reaching up to $\approx75\%$ \citep{GinzburgSyrovatskii1965} and has a redder spectrum. In case of XRPs, \mbox{Her X-1} and Swift~J0243.6+6124, synchrotron radiating jet was suggested to be the source of radio emission \citep{vdEijnden2018HerX1,vdEijnden2018J0243}. In case of the Be--X-ray binary, EXO 2030+375, exceptionally high PD ($\approx$16\%) was detected which was explained by the presence of non-spherical ferromagnetic grains in the Be disk that were aligned under the influence of the NS magnetic field \citep{Reig2014}. 
Thus, the observed intrinsic optical polarization in Be--XRP binaries can be a net result of the interplay between its different polarized components and their variability.

\subsection{Origin of the intrinsic optical polarization from \lsv} 
 The nature of the significant intrinsic optical polarization seen in \lsv can be attributed to one or more of the above mentioned components by studying its variability with time and its spectral dependence.
 
\subsubsection{Irradiation due to X-rays from the NS}

The X-ray light from the XRP can be reprocessed on the surface of the accretion disk around the NS or the circumstellar envelope of the Be star and thereby be observed as linearly polarized optical emission. In such cases, there is a time delay between the variations in X-rays and in optical bands.  
The semi-major axis of \lsv orbit inferred from the orbital period \citep[that is, 150 days as given in][]{Ferrigno2013} is $\approx 10^{13}$~cm.
The typical delay for light to travel across this distance is on the order of $\approx 300$~s and is greater than the pulsar period. 
Thus, any variability on the timescales of the pulsar period due to the X-ray source will be smeared out and hence remain indiscernible at distances close to the Be companion. 
Thus, the lack of significant variability in Fig.~\ref{fig:PDPA_pulsar} with respect to the pulsar period points towards outer accretion disk or the Be decretion disk as plausible sources of polarized emission in the optical bands.

If the optical linear polarization is coming from XRP irradiation, then, one can expect to see some change in both PD and PA over the orbital period as well. As the NS travels closer to (or further away from) the periastron, the accretion disk will be replenished (or depleted) and the observed polarization should react accordingly. Alternatively, the solid angle of irradiation onto the Be disk also varies with the orbital period, potentially altering the polarization. From Fig.~\ref{fig:PDPA_orbit}, it is not clear whether there exists any orbital variability. While strong peaks in PD and constant PA with phase are plausible in case of highly eccentric orbits, in order to conclude whether the variation is with respect to the orbital period or due to some other long term process, more data is required.

\subsubsection{Synchrotron emission from the XRP}

Another possible source of optical polarization is synchrotron emission from the XRP. Close to the XRP, the magnetic field ($B_{\rm mag}$) is too strong \citep[$\approx 10^{12}$~G; ][]{Tsygankov2012} to produce cyclo-synchrotron emission in the optical wavelengths.
As the dipole magnetic field strength falls $B_{\rm mag} \propto 1/r^3$ with distance from the XRP, and assuming that the optical emission is produced from a region where the magnetic field is $10^5$~G \citep[generally considered for optical synchrotron emission from the hot accretion flow in black hole X-ray binaries; ][]{Veledina2013}, the typical distances away from the NS where one can expect cyclo-synchrotron emission is $\approx 2 \times 10^8$~cm. This is around the magnetospheric radius of the XRP, which is the expected minimum distance from where the jet is launched in such systems \citep{vdEijnden2018J0243}. However, the intrinsic PD of \lsv does not have a red spectrum (Fig.~\ref{fig:PDPA_wavelength}a) expected from synchrotron emission. Furthermore, from X-ray polarimetry \citep{Doroshenko2023}, the minimum angles between magnetic dipole axis and the observer as the NS rotates changes from $60\degr$ (Northern pole) to $24\degr$ (Southern pole). Since the PA must be perpendicular to the magnetic field axis, it must also vary with the rotation period of the XRP which is not seen (Fig.~\ref{fig:PDPA_pulsar}). Thus, we conclude that the XRP synchrotron emission is not a significant source of optical polarization in \lsv.

\subsubsection{Be decretion disk}

From Fig.~\ref{fig:timeEvol}, we see that PD and PA do not show dramatic changes with time despite a change in X-ray flux of about two orders of magnitude. This suggests that accretion related effects are likely not the dominant mechanisms producing optical polarization. 
Instead, the long-term variability seen in the optical polarization of the source (Fig.~\ref{fig:timeEvol}, Tables~\ref{table:obs} and  \ref{table:t2test}) could be due to intrinsic changes in the Be decretion disk. Precession of the decretion disk or the formation and dissipation processes of the decretion disk, observed for \lsv through spectroscopic studies of its emission lines, reveal itself as variability on the timescale of tens of years \citep{Reig2005, yan2016long}. 
Evolution of Be stars and their decretion disks on such timescales is seen in a number of Be--binaries and manifests itself through super-orbital variability on the timescale of thousands of days \citep{Treiber2025}. 

The spectral dependence of the intrinsic polarization of \lsv shown in Fig.~\ref{fig:PDPA_wavelength}
 can be associated with what is expected from a plane-parallel scattering atmosphere. 
 It is likely that the optical polarization is originating from the decretion disk of the Be star rather than the accretion disk of the NS based on the time evolution of the optical polarization. A sole Be disk origin of the optical polarization implies that the decretion disk was not completely disrupted due to the outburst, which is otherwise theoretically plausible in such X-ray binaries \citep{rivinius2013classical}. This is in agreement with previous studies that suggest that \lsv continues its accretion even in the quiescent state \citep[][that is, the NS accretion disk is being replenished, plausibly by the Be disk matter]{reig1999discovery, LaPalombara2012}. 

 Considering the scenario that scattering in the Be disk is responsible for optical polarization, the PA can provide information on the orientation of the decretion disk. 
 The wavelength dependence of PA (Fig.~\ref{fig:PDPA_wavelength}b) implies warping of the decretion disk. This is conceivable when the inner and outer parts of the disk no longer remain co-planar post-outburst.
 Small changes in intrinsic optical PA (Table~\ref{table:intrinsic}) between epochs points towards a possibility of a precessing Be disk. Be--XRP sources have shown evidence of a precessing Be disk via a drift in the periastron phase of the NS following a giant outburst \citep{Huang_2025,Yongfeng2026}. Precessing Be disks could result in skipped Type-I outbursts over the next few orbits and/or result in outbursts mid-orbit. Activity from \lsv had been low since the giant outburst in 2023 until March 2026 (and subsequently in May and July 2026) when an increase in X-ray flux was detected \citep{LeStum2026ATel,Nakajima2026AtelMay,Usuki2026AtelJuly}. All of these provide strong evidence towards a precessing Be disk, which could also potentially be the source of optical polarization.

 In addition to rigid precession, Be disks have also been known to show signatures of global one-armed density oscillations through H$\alpha$ line profiles \citep[$V/R$ asymmetric variability; ][]{Hanuschik1995,Hummel1997}. The $V/R$ variability pattern has also been observed in \lsv \citep{Reig2005}. In this scenario, the small but significant PA evolution would reflect the slow rotation of an asymmetric over-density pattern in the Be disk. 

 Variability due to either stellar pulsations from the interior of the Be star or due to rotation of the Be star in \lsv have also been identified in \citet{reig2022fast} on the order of a few cycles per day. While such non-radial stellar pulsations could further influence the observed optical polarization, the interpretation of the same is limited by the prevailing observational data (Fig.~\ref{fig:PDPA_stellar}). 
 
 The obtained intrinsic optical PA in the range of $59\degr$--$68\degr$ is in close agreement with the PA of the pulse-independent component of X-ray polarization \citep[$\approx$70\degr; expected to originate from scattering in the outer accretion disk and wind][]{Doroshenko2023,Nitindala2025}, implying that the accretion disk lies along the decretion disk plane (especially the outer disk regions probed by the $R$ band). It is also possible that the constant component to X-ray polarization originates from scattering of XRP radiation in the Be decretion disk.

\section{Summary}\label{sec:summary}

We present optical ($BVR$) polarization measurements of the Be--XRP binary, \lsv, taken over three years, including the giant outburst in 2023. The observed PD of the source is $1\%$--$2\%$ with confidence levels over $10\sigma$ and is seen to be significantly evolving over timescales of $\approx$1000~d. 
We also checked the dependence of observed polarization with the orbital, stellar pulsation, and XRP rotation phase but find no strong evidence for any of them.

To find the polarization intrinsic to the source, we subtracted the interstellar polarization obtained from the weighted average optical polarimetric measurements of the field stars. The interstellar PD was found to be 2.5\%--3\% along the same direction as the observed polarization. This resulted in the intrinsic PD of \lsv being between $0.5\%$ and $1.7\%$ and the intrinsic PA being between $59\degr$--$68\degr$. 
Intrinsic PD of \lsv dropped from $B$ band towards $R$ band while the intrinsic PA shows a clear spectral dependence. 

We discuss various possible sources of optical polarization in the Be--XRP system. Considering both, the time and spectral evolution of polarization, scattering in the Be companion decretion disk best interprets the data. The data also points towards potential super-orbital variability on the timescale of thousands of days corresponding to a precessing Be disk, possibly with over-density oscillations.

\begin{acknowledgements}
This study was  supported by the Research Council of Finland grants 355672 and 372881 and Centre of Excellence in Neutron-Star Physics (grant 374064).
APN acknowledges support from the Magnus Ehrnrooth Foundation. 
Nordita is supported in part by NordForsk. 
The DIPol-2 and DIPol-UF polarimeters were built in cooperation between the University of Turku, Finland, and the Kiepenheuer Institut fuer Sonnenphysik, Germany, with support from the Leibniz Association grant SAW-2011-KIS-7 and the ERC Advanced Grant HotMol, ID: 291659. 
The data collected using DIPol-UF were based on observations made with the Nordic Optical Telescope, owned in collaboration by the University of Turku and Aarhus University, and operated jointly by Aarhus University, the University of Turku and the University of Oslo, representing Denmark, Finland and Norway, the University of Iceland and Stockholm University at the Observatorio del Roque de los Muchachos, La Palma, Spain, of the Instituto de Astrofisica de Canarias. The NOT data were obtained under program ID P72-025.
This work has made use of data from the European Space Agency (ESA) mission {\it Gaia} (\url{https://www.cosmos.esa.int/gaia}), processed by the {\it Gaia} Data Processing and Analysis Consortium (DPAC, \url{https://www.cosmos.esa.int/web/gaia/dpac/consortium}). 
Funding for the DPAC has been provided by national institutions, in particular the institutions participating in the {\it Gaia} Multilateral Agreement. 
This research has made use of \textit{"Aladin sky atlas"} developed at CDS, Strasbourg Observatory, France  \citep{Aladinv3}.
VK acknowledges support from Turun Yliopistosäätiö.
\end{acknowledgements}

\bibliographystyle{yahapj} % style aa.bst
\bibliography{references.bib} % your references Yourfile.bib

\end{document}